\documentclass[journal]{IEEEtran}
\ifCLASSINFOpdf

\else
\fi

\usepackage{graphicx,amsmath,amssymb,cite}
\usepackage{subfigure}
\usepackage{color,xcolor}
\usepackage{amsfonts}
\usepackage{amsmath}
\usepackage{graphicx}
\usepackage{amssymb}
\usepackage{stfloats}
\usepackage{mathrsfs}
\usepackage{algorithm}
\usepackage{algorithmic}
\usepackage{graphicx}  %插入的宏包
\usepackage{epstopdf}
\usepackage{amssymb}
\usepackage{amsfonts}
\usepackage{amsmath}
\usepackage{algorithm}
\usepackage{algorithmic}
\usepackage{subeqnarray}
\usepackage{cases}
\usepackage{soul}
\usepackage{makecell}%表格中内容换行命令

\def\BibTeX{{\rm B\kern-.05em{\sc i\kern-.025em b}\kern-.08em
    T\kern-.1667em\lower.7ex\hbox{E}\kern-.125emX}}
\begin{document}

%\title{An efficient deep learning network structure for secure spatial modulation  receiver }
\title{Machine-learning-aided Massive Hybrid Analog and Digital MIMO DOA Estimation for Future Wireless Networks}

\author{Xinyi Zhao, Baihua Shi, Yiwen Chen, Xichao Zhan, Feng Shu, Wenlong Cai, Mengxing Huang, Qijuan Jie, Yifang Li, Jiangzhou Wang, \emph{ Fellow, IEEE} and Xiaohu You, \emph{ Fellow, IEEE}

\thanks{This work was supported in part by the National Natural Science Foundation of China (Nos.U22A2002, and 62071234), the Major Science and Technology plan of Hainan Province under Grant ZDKJ2021022, and the Scientific Research Fund Project of Hainan University under Grant KYQD(ZR)-21008. (Corresponding authors: Mengxing Huang, Wenlong Cai and Feng Shu).}
\thanks{Xinyi Zhao is with the School of Mathematics and Statistics, Nanjing University of Science and Technology, Nanjing, 210094, China.}
\thanks{Feng Shu, Yifang Li, and Baihua Shi is with the School of Electronic and Optical Engineering, Nanjing University of Science and Technology, Nanjing, 210094, China.}
\thanks{Feng Shu, Yiwen Chen, Xichao Zhan, Qijuan Jie, and Mengxing Huang are with the School of Information and Communication Engineering, Hainan University, Haikou, 570228, China. (Email: shufeng0101@163.com).}
\thanks{Wenlong  Cai is  with the National Key Laboratory of Science and Technology on Aerospace Intelligence Control, Beijing Aerospace Automatic Control Institute, Beijing 100854, China. (Email: {caiwenlon}@buaa.edu.cn).}
\thanks{Jiangzhou Wang is with the School of Engineering, University of Kent, Canterbury CT27NT, U.K. (Email: j.z.wang@kent.ac.uk)}
\thanks{Xiaohu You is with the School of Southeast University, Nanjing, 211189, China. (Email: xhyu@seu.edu.cn).}
}
\maketitle

\begin{abstract}
	
 %\textcolor{blue}{The exploration of green wireless communication technologies is critical to achieving the sustainable development for many emerging services within the strict constraints of data rates and latency}.
 Due to a high spatial angle resolution and low circuit cost of massive hybrid analog and digital (HAD) multiple-input multiple-output (MIMO), it is viewed as a valuable green communication technology for future wireless networks. Combining a massive HAD-MIMO with direction of arrival (DOA) will provide a high-precision even ultra-high-precision DOA measurement performance approaching the fully-digital (FD) MIMO. However, phase ambiguity is a challenge issue for a massive HAD-MIMO DOA estimation. In this paper,  we review  three aspects: detection, estimation, and Cramer-Rao lower bound (CRLB) with low-resolution ADCs at receiver. First, a multi-layer-neural-network (MLNN) detector  is proposed to infer the existence of passive emitters. Then, a two-layer HAD (TLHAD) MIMO structure is proposed to eliminate phase ambiguity using only one-snapshot. Simulation results show that the proposed MLNN detector  is much better than both the existing generalized likelihood ratio test (GRLT) and the ratio of maximum eigen-value (Max-EV) to minimum eigen-value (R-MaxEV-MinEV) in terms of detection probability. Additionally, the proposed TLHAD structure can achieve the corresponding CRLB using single snapshot.
\end{abstract}

\section{Introduction}

The position of a target of interest can be inferred by using its emitted signal
measured at an array of spatially separated nodes, where the positions of nodes are known.  Actually, source localization has been become one of the crucial issues in many research fields such as robot, mobile communications, radar, sonar,  wireless sensor networks, satellite communications, human-computer interaction, and marine communications \cite{Handbook-19,HuangDL0223TVT,shuDM2021twc,shuDM2020network}. Source localization techniques  fall into the main five categories:  direction of arrival (DOA) in \cite{T.Engin-09}, received signal strength indicator (RSSI) in \cite{Wang-sichun-13} , time of arrival (TOA), time difference of arrival (TDOA), and frequency difference of arrival (FDOA). Different from other methods, RSSI may  work in both line-of-sight (LOS) and Non-LOS (NLOS) environments and is insensitive to the availability of multi-paths. In particular,  the remaining four techniques are more suitable for the LOS scenario. In the presence of multipaths, their localization accuracies will be degraded substantially.

%In recent years, with the emergences of the massive antenna array structure, the DOA estimation, as a traditional field, has attracted an increasingly attentions from academia and industy  due to its ultra-high-angle-resolution\cite{SF-HAD-DOA-18,HuDie}.
With the deep integration of artificial intelligence, mobile communication and other technologies (ICDT), some emerging services such as metaverse and holographic communication have higher demands for end-to-end information processing rates and latency, enabling the integrated sensing and communication (ISAC) one of the leading trends of 6G technology. In recent years, with the emergences of the massive antenna array structure, DOA estimation, as a traditional field, can provide a ultra-high accurate desired signal direction for beamforming and tracking, and achieve a higher signal-to-noise ratio at the receiver with less transmit power compare to conventional small MIMO receiver\cite{SF-HAD-DOA-18,HuDie,wenDOA2023TWC}. As one of the key technologies in the beyond fifth generation (B5G) and sixth generation(6G) mobile communication systems \cite{gui2020ISAC}, DOA estimation will gradually serve in the construction of ISAC , and provide a high energy-efficient green beamforming communication. As the number of antennas tends to large-scale or utra-large-scale, it is possible to achieve a localization accuracy of centimeter via DOA. Unluckily, this  leads to a  high circuit cost like massive antennas, analog-to-digital convertors (ADCs), and radio frequency links, etc.  The corresponding  computational complexity is significantly increased. Hybrid analog and digital (HAD) MIMO array becomes a natural choice, which can dramatically reduce energy consumption and circuit costs and enable a reliable green wireless communication technology to support future sustainable development of applications such as metaverse or web3.0 \cite{SF-HAD-DOA-18,SBH-20}. Three high-performance DOA estimation methods were proposed in \cite{SF-HAD-DOA-18} for eliminating phase ambiguity of massive hybrid MIMO systems, and the Cramer-Rao lower bound (CRLB) was derived. In \cite{SBH-20}, a fast ambiguous phase elimination method was proposed to find the true direction using only two snapshots at the cost of a slight performance loss. The DOA estimation problem of sparse array design with non-circular (NC) signals was investigated in \cite{Zhang-xiaofei-21}. Furthermore, in \cite{Meng-Xiangming-18}, the generalized sparse Bayesian learning algorithm was integrated into the 1-bit DOA estimation. Adopting massive MIMO receive array with low-resolution ADC, demonstrated a new DOA estimation scheme in \cite{SBH-ADC}.

Due to the adoption of eigenvalue decomposition, the computational complexities of the above matrix-decomposition-based DOA estimators are high. A novel deep-learning framework of achieving super-resolution channel estimation and DOA estimation was proposed in massive MIMO systems in \cite{Huang-hongji-18}. In \cite{HuDie}, a low-complexity deep-learning based DOA estimator with uniform circular arrays (UCA) was proposed for the massive MIMO systems with much lower computational complexity than conventional maximum likelihood (ML) method. Afterwards, in \cite{Zhuang-zhihong-20}, a novel estimating signal parameter via rotational invariance techniques (ESPRIT) method and a machine learning framework were proposed to improve the accuracy of DOA measurements for the HAD structure. Due to its ultra-high precision of angles using a massive or ultra-massive MIMO, the positioning based on the angle of arrival (AOA) can reach a high-performance localization accuracy. For exmaple, in \cite{SBH-zongshu}, a geometric center AOA localization method was proposed utilize single base station, which can achieve the CRLB. By estimating the DOA was given to achieve the associated CRLB by forming a polygon of intersecting multiple estimated DOA lines. The final source position is the center of this polygon. Aiming at the indoor localization, the paper \cite{Li-Yiwen-20-5g} proposed a fifth generation (5G) signal based localization method by employing the estimates of the AoA and the time of flight, which can achieve a centimeter-level localization accuracy. In \cite{WangYue-AOA-TDOA}, the authors presented an improved polarity representation to uniformly use the AOA the localization of a signal source, regardless of whether it is a near or far.
\begin{figure*}[htbp]
\centering
\includegraphics[width=0.8\textwidth,height=0.22\textheight]{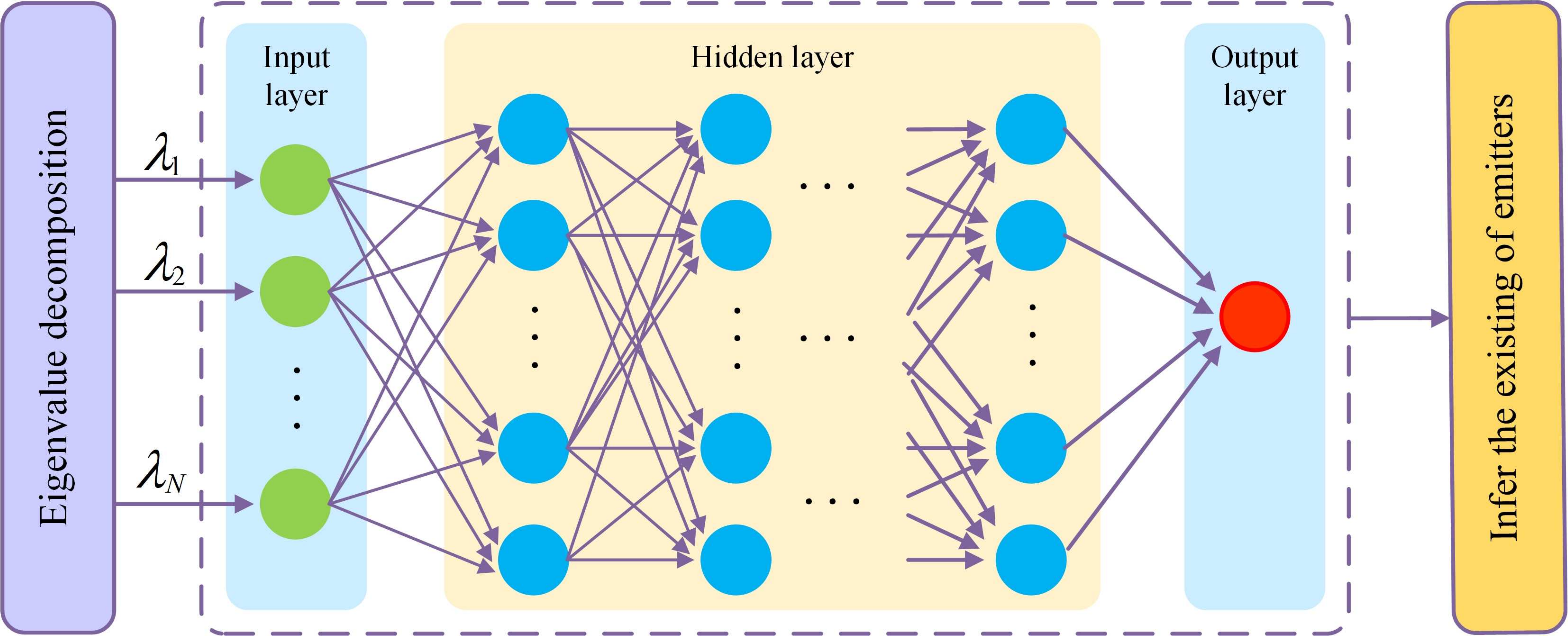}\\
\caption{Proposed MLNN detector}\label{fig_DNN_system.eps}
\end{figure*}

\section{Proposed multi-layer neural network detector for multi-emitter}
Since the turn of the twentieth century, direction-finding using MIMO receive array has been investigated extensively so far. The  research efforts focused mainly on DOA measurement methods, CRLB, and array calibration etc. To the best of our knowledge, there are few research works on passive emitter detection. Similar to RADAR, an active way of finding direction and range, it is also mandatory for passive direction-finding to infer whether the emitter exists or not before performing DOA estimation operation. For example, when there exists no emitter, a DOA measurement operation is directly conducted. Obviously, this will output the direction angle of a virtual emitter,  result in a form of false alarming, and waste the computational amount at MIMO receiver.

To achieve a high detection performance of emitter and trigger the next step: DOA measurements, a high-performance detector was proposed to infer the existence of multi-emitter from the eigen-space of sample covariance matrix of receive signal vector \cite{jie-qi-juan-MIMO}. Here, the sampling covariance of receive signal vector was first computed, and its EVD was performed to extract all its eigenvalues. The test statistic is defined, as the ratio of maximum eigen-value (Max-EV) to minimum eigen-value, called R-MaxEV-MinEV. Their closed-form expressions were presented and the corresponding detection performance was given. As shown in \cite{jie-qi-juan-MIMO}, the proposed  R-MaxEV-NV method  performs much better than the traditional generalized likelihood ratio test (GLRT) method with a fixed false alarm probability in terms of receiver operating characteristic curve (ROC).

To improve the ROC performance, a multi-input single-output (MISO) binary classification of multi-layer-neural-network (MLNN) is proposed in this article. Here, all the eigenvalues of the covariance matrix of the received signal is used as the input signal of the MLNN. A gradient descent algorithm is used to train the neural network. The output of the neural network is a value in the interval [0,1] and an appropriate threshold is determined by the ROC. When the output of the network is less than the threshold, it means that the emitter does not exist and returns to the previous step, and when the output of the network is greater than the threshold, it means that the emitter exists and continues speculation. The training set is constructed by Monte Carlo simulation, and only one neuron is needed for the output layer.

Fig.~\ref{fig_DNN_system.eps} sketches the block diagram of the passive emitter detection system based on MLNN. Firstly, a set of sample vectors is collected within  one snapshot, and  the corresponding covariance matrix is estimated. Secondly, an eigen-decomposition (EVD) is made on the estimated covariance matrix  to obtain all its eigenvalues, which are used as the input signal of MLNN.

The total MLNN training process is divided into three stages. In the first stage,  the training data is used to  learn the MLNN activation function, and  the activation functions of each layer is optimized, including Sigmoid, Tanh, Rectified linear unit, etc., to explore the optimal activation function configuration. In the second stage, the appropriate network depth and the number of neural units in each hidden layer are attained. The main purpose of the hidden layer is to extract channel features, such as clutter and noise co-square matrix. Finally, after determining the network depth and the number of neurons per layer, the random gradient method is applied to minimize the variance and obtain all the weighted coefficients of the MLNN, in order to avoid under fitting and over fitting, the amount of data in the training set is 5-10 times the total number of weighted coefficients. Finally, the above MLNN is directly applied to the massive MIMO receiver to test its false alarm and detection probability performances.

Below,  the practical detection performance simulation is conducted to evaluate the detection probability of the three detection methods. Without loss of generality, system parameters are chosen as follows: $N=64$, SNR=-20dB, $L=200$.

\begin{figure}[h]
\centering
\includegraphics[width=0.48\textwidth]{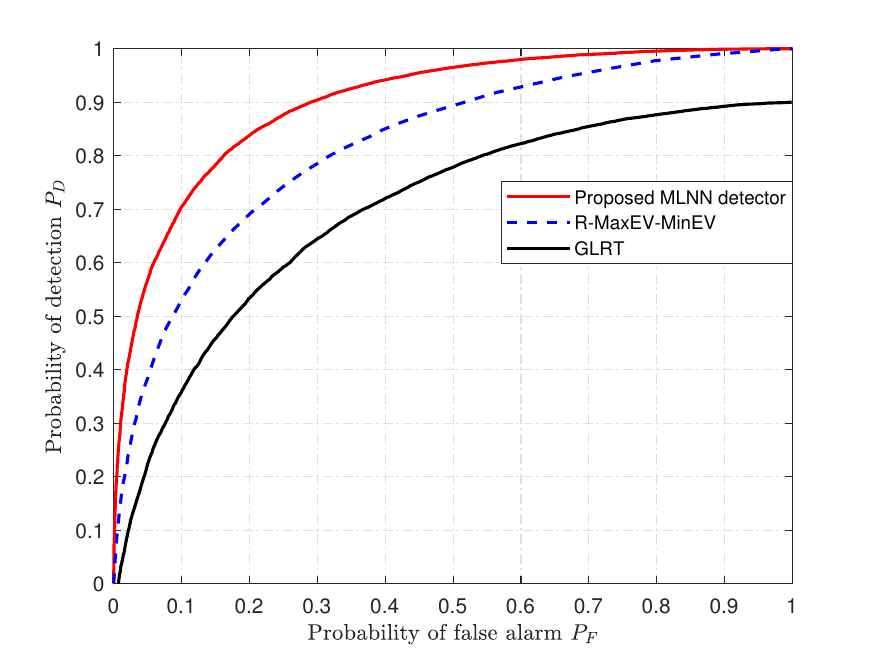}\\
\caption{ROCs under three different detection methods}\label{fig_MLNN_ROC}
\end{figure}

\begin{figure*}[httb]
\centering
\includegraphics[width=0.8\textwidth,height=0.24\textheight]{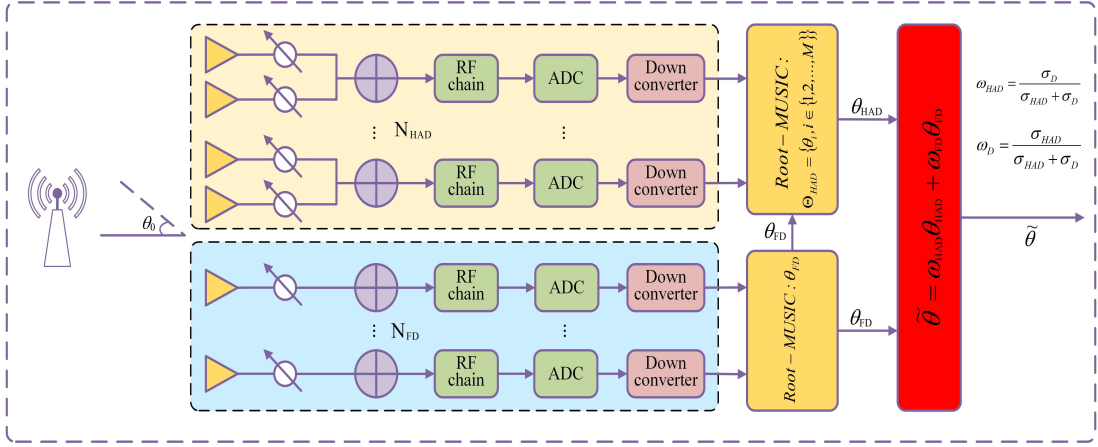}
\centering
\caption{Proposed two-layer HAD structure}\label{double-layer HAD structure_fig1}
\end{figure*}

Fig.~\ref{fig_MLNN_ROC} plots the ROCs for the proposed MLNN detector with conventional GLRT and R-MaxEV-MinEV as performance benchmarks. It can be seen from Fig.~\ref{fig_MLNN_ROC} that the detection performance of the proposed MLNN detector for multi-emitter is much better than existing GLRT for a fixed false alarm probability (FAP). Additionally, the R-MaxEV-MinEV is slightly better than GRLT in terms of detection probability given a small FAP.

\section{Proposed two-layer HAD structure}

A HAD MIMO receiver is  very suitable for DOA measurements due to its low circuit cost and power consumption  at the expense of some performance loss, especially for large-scale or ultra-large-scale scenario \cite{SF-HAD-DOA-18,HuDie}. The major drawback of HAD is that there exists phase ambiguity. This means  the DOA measurement method using HAD is made up of two  steps:  forming a set of candidate solutions by conventional DOA estimators like Root-MUISC  and eliminating  spurious solutions in such a set to find the true direction angle in \cite{SF-HAD-DOA-18,SBH-20}. In \cite{SF-HAD-DOA-18}, three high-performance DOA measurement methods were proposed  and the corresponding  CRLB was also derived. Among them, the best one is the HAD-Root-MUISC, which can achieve the hybrid CRLB with a lower-complexity than others. However, they require $M+1$ snapshots to implement one-time DOA estimate, where $M$  is the number of antennas per subarray. This results in a large processing delay. To reduce this delay, an improved fast HAD-Root-MUSIC method was proposed in \cite{SBH-20}, where the analogy phase alignment was divided into multiple sub-groups, and the analog beamforming of the sub-arrays of each sub-group was arranged at a candidate angle in a time block to eliminate the ambiguity of the phase. Compared with \cite{SF-HAD-DOA-18}, it needs only two snapshots to find the true direction angle of emitter at the cost of a little performance loss.

Does it exist an one-snapshot DOA measurement method ?  To address this problem, a two-layer HAD receive structure is proposed  in Fig.\ref{double-layer HAD structure_fig1}. The total receiver consists of three parts: a HAD on the left-top corner, a fully-digital (FD) on the left-bottom-left corner, and a combiner on the right side. The first part HAD is to generate a set of candidate solutions by using some traditional methods like Root-MUSIC. The second part FD  is to estimate the true solution, which will be adopted to remove the spurious solutions in candidate set.  The final part combiner is to combine the two true solutions to output an improved solutions. How to choose the weight coefficients of this combiner will affect the resulting performance of the two-layer structure. In order to achieve a good performance, they are chosen to be proportional to the corresponding CRLBs.

\begin{figure}[htbp]
\centering
\includegraphics[width=0.48\textwidth]{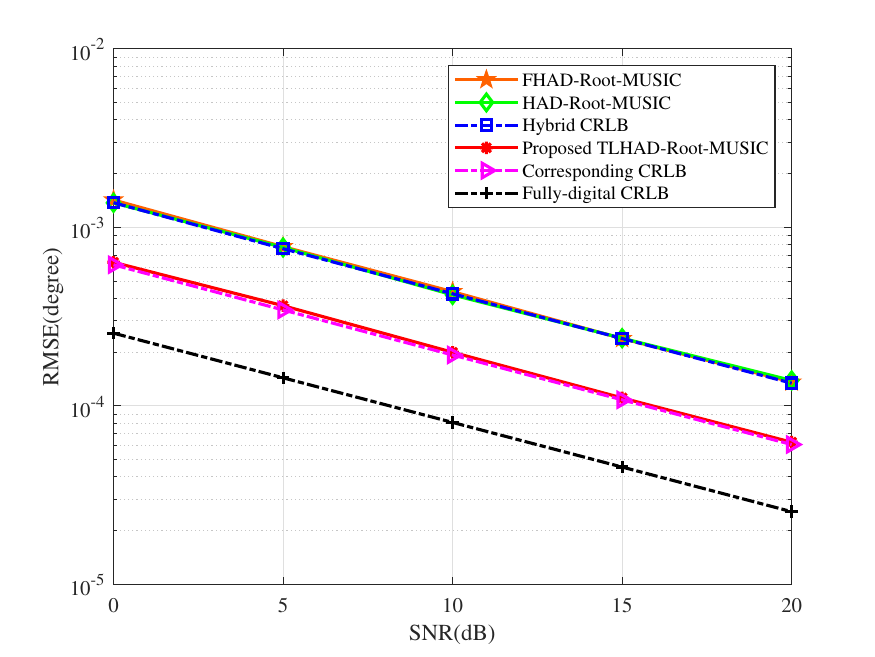}\\
\caption{RMSE versus SNR of the proposed method}\label{RMSE-SNR}
\end{figure}

Fig. \ref{RMSE-SNR} plots  the curves of RMSE versus SNR of the proposed method with existing method and CRLBs as performance benchmarks. From this figure, it is seen that the proposed method can achieve the corresponding CRLB and performs better than existing methods HAD-Root-MUSIC and FHAD-Root-MUSIC  in \cite{SF-HAD-DOA-18,SBH-20}.  This is mainly due to the fact that the proposed structure using a FD sub-structure to replace the corresponding HAD substructure. The proposed structure actually is a mixture of FD and HAD. Obviously, increasing the portion of FD in the total structure will improve the RMSE performance of estimating.

\begin{figure}[htbp]
\centering
\includegraphics[width=0.48\textwidth]{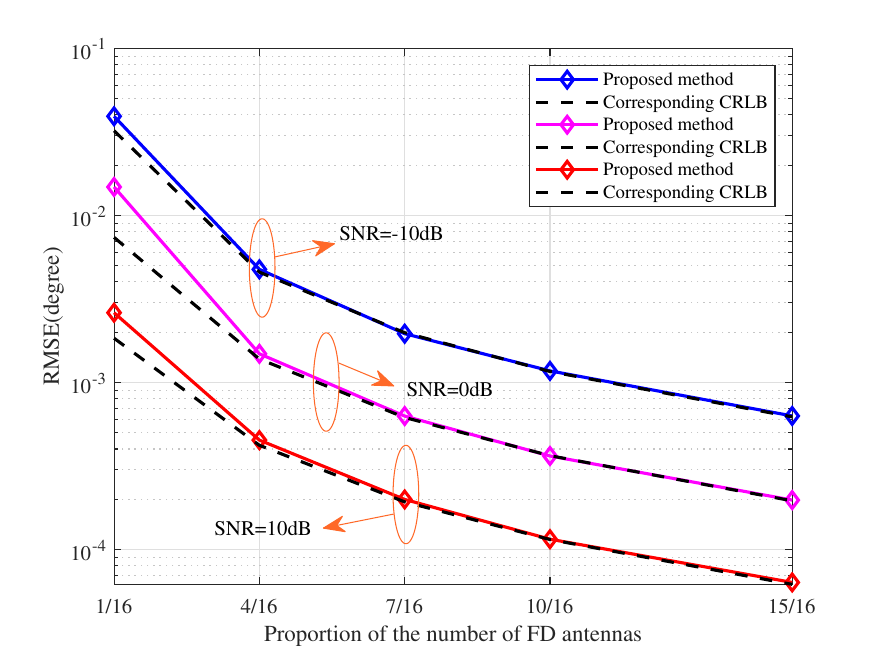}\\
\caption{RMSE versus proportion of FD of the proposed method}\label{RMSE-numbers}
\end{figure}

To evaluate the impact, Fig. \ref{RMSE-numbers} plots  the curves of RMSE versus the proportion of FD of  the proposed structure for three different SNRs (-10dB, 0dB, and 10dB) with the corresponding CRLBs as performance benchmarks. Observing this figure, we find that the proposed structure still can achieve the two-layer HAD CRLB for $\eta\ge 25\%$. It means that the appropriate proportion of FD can be selected for the performance requirements of different scenarios.

\section{Performance analysis of HAD-MIMO DOA measurments with low-resolution ADCs}
As the number of antennas goes to large-scale in massive MIMO receiver, the number of ADCs increases accordingly. High-resolution ADCs means high cost and energy consumption. Adopting low-resolution ADCs is a promising solution to reduce the circuit cost and power consumption in future wireless communications. Low-resolution structure has attracted many attentions. By resorting to the compressive sensing, the DOA estimation with 1-bit ADCs was solved as a sparse recovery problem.  The multiple signal classification (MUSIC) method can be directly utilized to perform DOA estimation with 1-bit ADCs. However, to the best of our knowledge, the  performance loss was not derived and analyzed due to the use of low-resolution ADCs. Thus, in \cite{SBH-ADC}, the authors proposed a performance loss factor to evaluate the performance loss in massive MIMO systems with low-resolution ADCs. By employing the additive quantization noise mode (AQNM), the closed-form CRLB expression was derived. Based on that, a new performance loss factor was defined. It is shown that the performance loss factor is related to SNR and number of quantization bits. Additionally, it is also verified that all subspace-based methods can be used in low-resolution ADC architecture without any modification.

\begin{figure}[htbp]
  \centering
  \includegraphics[width=0.48\textwidth]{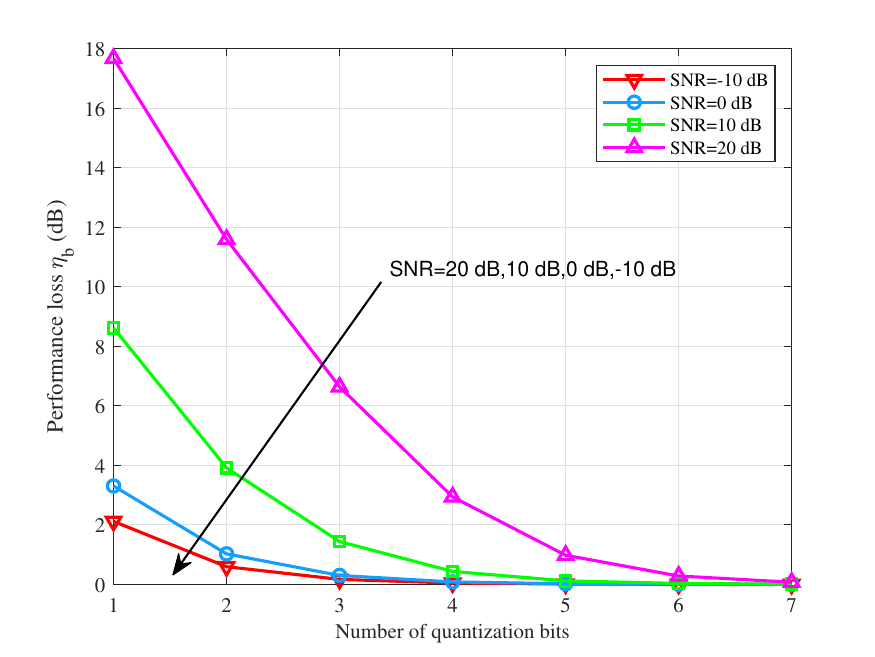}\\
  \caption{Performance loss versus number of quantization bits with $N=32$ and $M=128$ for different SNRs.}\label{PL_bit}
\end{figure}

Fig. \ref{PL_bit} plots the performance loss versus the number of quantization bits for different SNRs. It is obvious that performance loss decreases as the number of quantization bits increases. In addition, the performance loss is a decreasing function of  SNR. If we set $1$ dB loss as an acceptable loss, 2-bit ADCs are suitable in the low SNR region. However, as the SNR increases, it is better to adopt 3-bit ADCs in the medium SNR region. It is recommended to adopt $4\thicksim5$ bit ADCs in the high SNR region. Furthermore, ADCs with $b>5$ only achieve a trivial  performance gain over those with lower-resolution.
%\begin{figure}[t]
%  \centering
%  \includegraphics[width=0.48\textwidth]{fig_RMSE_SNR_RandE.eps}\\
%  \caption{RMSE over SNR for Root-MUSIC and ESPRIT.}\label{RandE}
%\end{figure}

%Fig. \ref{RandE} demonstrates the RMSE versus SNR for two classic subspace-based DOA methods: Root-MUSIC and  estimation of signal parameters via rotational invariance technique (ESPRIT). Obviously, both Root-MUSIC and ESPRIT can achieve the corresponding CRLBs in the low SNR region. Thus, simulated results proved that subspace-based method can be used in low-resolution ADC structure without any modification. As SNR increases, there are some gaps between simulated curves and analytical curves. Because AQNM is not accurate enough at high SNRs. Moreover, root-MUSIC can achieve the CRLB, but ESPRIT failed. This results from  essences of algorithms. Root-MUSIC has higher performance and ESPRIT has lower computational complexity.

\section{Open Challenging Problems}
However, there are still several important open problems arising in DOA estimation field using massive HAD-MIMO  summarized as follows:
\begin{enumerate}
\item As the number $N$ of antennas tends to large-scale or ultra-large-scale, the computational complexities of emitter detections and DOA measurements increase as a function of $N^3$. For example, $N=1000$, its complexity reaches up to $10^9$  FLOPs. Thus, there is a large demand for low-complexity emitter detectors and DOA estimators in the case of large-scale.
\item For multi-emitter scenario,  only detection is not sufficient. Obviously, a further task is to infer the number of passive emitters using deep learning methods. This will be very helpful for aiding the next step DOA measurement.
\item DOA estimation using ultra-massive MIMO systems could achieve an ultra-high precision of angles, which could pave the way to the AOA localization. How to use multiple estimated AOAs to realize a high-performance and low-complexity localization from geometric aspect? There are two ways:  using multiple distributed massive receive MIMO arrays or dividing a ultra-massive array into several subarrays. Each array or subarray makes an independent DOA measurement, and the measured DOAs are used to form several intersection points. The key is how  those intersection points  are explored to form a high-precision geometric intersection localization method of achieving the corresponding CRLB.
\end{enumerate}

\section{Conclusions}
In this article, a MLNN-based detector was proposed to improve the emitter detection performance. Subsequently, a DOA estimation algorithm with a TLHAD structure was proposed, which achieves the elimination of phase ambiguity in a single snapshot and makes a significant reduction on a DOA measurement time delay. Furthermore, a low-cost framework of DOA estimation combining low-resolution ADCs and large-scale HAD MIMO system was also reviewed, which strikes a good balance between performance and circuit cost. Finally, potential challenges of DOA estimation for massive or utra-massive MIMO systems were discussed, and several new open important issues were presented. Thus, considering its low-cost-circuit and high performance, massive HAD DOA measurement may become a promising green communication technology for many future applications like metaverse, web 3.0 and the ISAC of beyond 5G/6G, etc.
%find its wide future applications  like serving  the ISAC of beyond 5G/6G.

% bmeps -c fig4_DNN_system.jpg fig4_DNN_system.eps(convert a to b)

\ifCLASSOPTIONcaptionsoff
  \newpage
\fi
\bibliography{IEEEfull,reference}
\bibliographystyle{IEEEtran}

\end{document}